\documentclass[epj]{svjour}
\usepackage{amsmath,bm,amssymb,graphicx,psfrag}

\begin{document}

\title{Fracture and Friction: Stick-Slip Motion}

\author{Efim A. Brener\inst{1} \and 
S. V. Malinin\inst{1,2}\thanks{\emph{Present address:} 
Institut f\"ur Theoretische Physik, 
Universit\"at zu K\"oln, 50937 K\"oln, Germany} \and 
V. I. Marchenko\inst{3}
}
\institute{Institut f\"ur Festk\"orperforschung, Forschungszentrum
J\"ulich \and L.D. Landau Institute for Theoretical Physics, RAS,
119334, Kosygin str. 2, Moscow, Russia \and
P.L. Kapitza Institute for Physical Problems, RAS,
119334, Kosygin str. 2, Moscow, Russia}

\date{\today}

\abstract{
We discuss  the  stick-slip
motion of an elastic block sliding along a rigid substrate.
We argue that for a given external shear stress this system shows
a discontinuous nonequilibrium transition from a uniform stick state
to uniform sliding at some critical stress which is nothing but 
the Griffith threshold for crack propagation. 
An inhomogeneous mode of sliding occurs, when the driving velocity 
is prescribed instead of the external stress. A transition to
homogeneous sliding  occurs at a critical velocity, 
which is related to the critical stress. We solve the elastic
problem for a steady-state motion of a periodic stick-slip pattern
and derive equations of motion for the tip and resticking end 
of the slip pulses. In the slip regions we use the linear viscous 
friction law and do not assume any intrinsic instabilities even 
at small sliding velocities. 
We find that, as in many other pattern forming system, 
the steady-state analysis
itself does not  select uniquely all the internal parameters
of the pattern, especially the primary wavelength. Using
some  plausible  analogy to  first order phase transitions
we discuss a ``soft'' selection mechanism. This allows to estimate
 internal parameters such as crack velocities, primary wavelength and 
 relative fraction of the slip phase as function of the driving
velocity.  The relevance of our results to recent experiments is
discussed.
%
}

\maketitle

\section{Introduction}
The nature of sliding friction and especially of inhomogeneous modes of 
sliding is a fundamental physical problem of  prime practical importance 
\cite {Persson2000}. Studying the friction 
of rubber along a smooth glass substrate,
Schallamach \cite{Schallamach71} observed an inhomogeneous mode of 
sliding. Under some conditions homogeneous sliding
becomes unstable and a quasiperiodic pattern of detached zones is formed.    
Recent experimental observations of Rubio and
Galeano \cite{Rubio94},  Baumberger, Caroli, and Ronsin
\cite{Baumberger2002,Baumberger2003}, on the frictional motion of sheared gels
sliding along a glass surface also indicate the existence of self-healing
pulses and inhomogeneous modes of sliding. 
A regime of periodic
stick-slip has been observed in a limited range of small
shear rates \cite{Baumberger2002,Baumberger2003}. It bifurcates towards
stationary sliding at some critical driving velocity. The slip
pulses traverse the sample with a velocity much larger than the
driving velocity but still much smaller than the  speed of sound.
The importance of the slow crack-like fronts for the onset of the frictional 
slip was also stressed in \cite{Rubinstein04}.

Slip pulses in gels seem to be very different from Schallamach
waves and ``brittle'' pulses studied by Gerde and Marder
\cite{Gerde01} since no observable opening  occurs at the interface.
In this respect, they are more comparable to
self-healing cracks, suggested by 
Heaton \cite{Heaton90}
 in the
context of seismic events.

Stick-slip motion with the transition to sliding above some critical velocity
was also observed in studies of friction 
in an ultrathin layer of lubricant between two atomically flat surfaces
\cite{Israelachvili93}. This behavior was attributed to the 
confinement-induced freezing of the lubricant and its melting due to 
shear stress.  

Recent investigations (see, for example, \cite{Ranjith01} and
references therein) point towards an essential importance of the
underlying friction law in the slip state. It has been proved that
the simple Coulomb friction leads to the so-called
``ill-posedness'' of the linear stability problem for 
small inhomogeneous perturbations of the stress and strain fields
in a sliding mode \cite{Ranjith01}. Moreover, Caroli
\cite{Caroli2000} has shown that the existence of slow, periodic
slip pulses is incompatible with the Coulomb friction law.

In our previous paper \cite{Brener02} we developed a conceptually simple model
which  is compatible with an inhomogeneous mode of sliding,  
the existence of slowly moving slip pulses and the ``shear melting'' phenomenon.
We discussed  the propagation of a shear 
crack (Mode II crack) along the interface
between two dissimilar materials. The crack edge separates two states
of the interface, ``stick'' and ``slip''. We assumed that the
interface is flat with a strong adhesion contact.  
In the presence of roughness, the assumption of 
strong adhesion and full contact at the interface 
is presumably only reasonable  for  ``soft''
materials with a relatively small shear modulus. Gels are clearly
materials of this type.

In the slip region we assumed a simple linear viscous  friction law,
namely, that the shear stress is proportional to the sliding velocity.
This, from the theoretical point of view strongly motivated law,  
is  usually  not discussed in the literature since it
does not lead to the so-called static friction phenomenon observed
experimentally. However,  in our description  static 
friction appears in a natural way  as the usual Griffith
threshold for crack propagation. The important point is that, 
before the system goes  into a sliding mode, a shear crack should
traverse the sample. This requires a finite shear stress since the
stick state of the interface is energetically more favorable.

We note that the above mentioned  interface properties in our model
are fully coupled to 
the bulk elasticity of a gel leading to a rich spectrum of phenomena
where two intriguing problems, crack propagation and interface 
friction, come together.

In the present paper we attempt to find a solution of this model 
which represents a periodic pattern of stick-slip motion.
The paper is organized as follows:
 
In the next section we formulate 
the problem of the periodic pattern which appears in an elastic solid block
sliding on a flat rigid substrate (see Fig. \ref{stickslip}). 
We formulate the boundary conditions 
for this elastic problem and introduce the thermodynamics of two states 
of the interface: stick and slip. 
Each slip pulse can be considered as a mode II crack which  
is bounded by two crack edges: the pulse tip and 
the resticking end. In order to formulate equations of motion for such cracks 
we have to solve the elastic problem and calculate the energy flux into the
crack edges. 

In section \ref{sec:3} we solve the elastic 
problem in two limiting cases: the height of the block is much larger than 
the lateral scale of the pattern and  the opposite limit.   
This section is rather technical but eventually provides expressions for the 
energy flux into the crack tips. It turns out that, as in many pattern forming 
systems (for example, directional solidification or eutectic growth 
\cite{Kurz2001}), the steady-state description itself does not allow to select 
the primary wavelength of the pattern. We will see that the degeneracy of our 
system is even higher than in crystal growth problems. 

In section \ref{sec:4} we 
give some plausible arguments based on the analogy to the first order  
phase transitions which lead to the ``soft'' selection of the periodic 
pattern. We predict a regime of stick-slip motion in a limited range of small
shearing rates and a transition towards stationary sliding at some critical
driving velocity. 
In section \ref{sec:5} we discuss our results and their possible relevance to existing
experimental observations. 
Technical details are presented in the Appendix.

\section{Formulation of the problem}

\begin{figure}[h]
\includegraphics[width=1\linewidth]{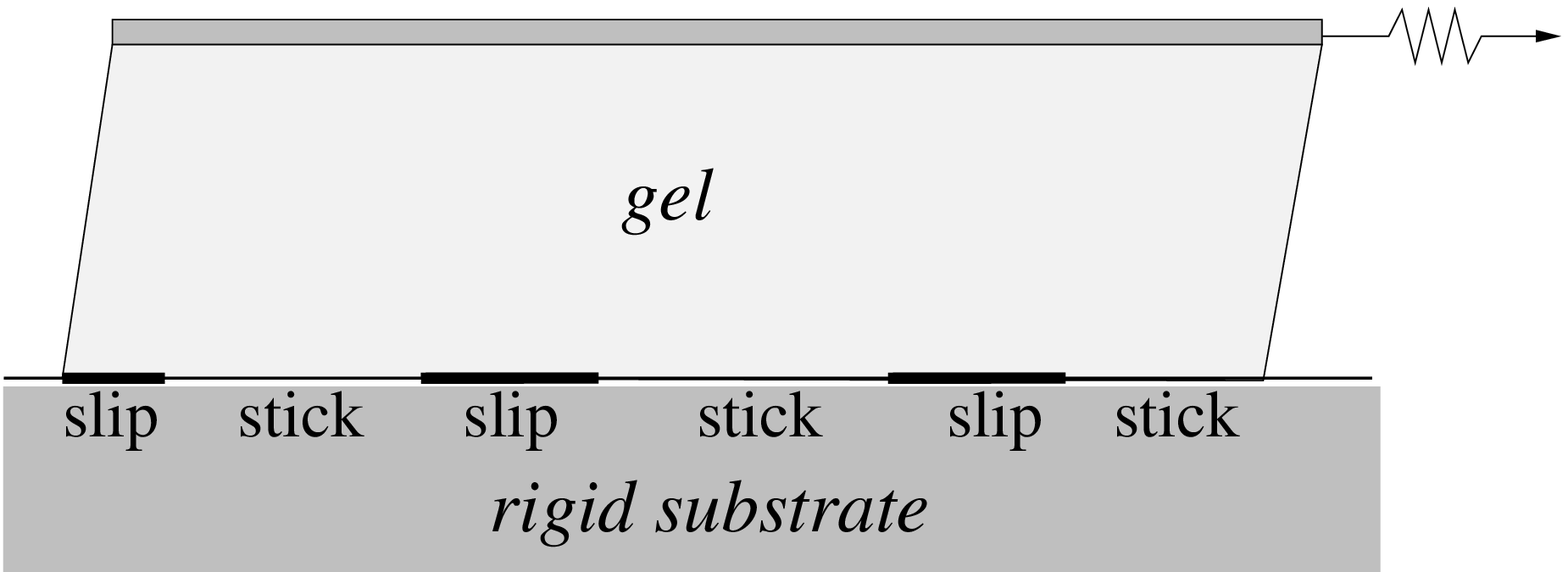}
\includegraphics[width=1\linewidth]{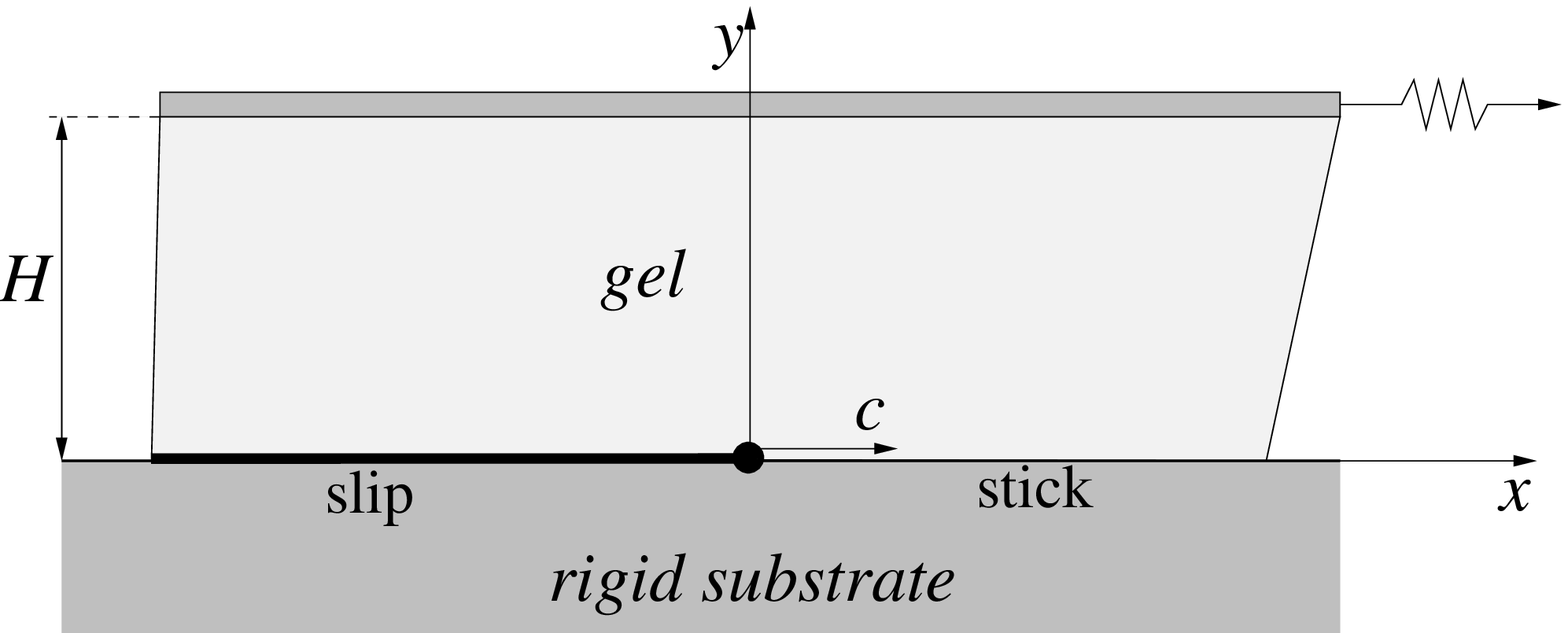}
\caption{(top) an elastic body sliding on a rigid substrate;
(bottom) motion of a single crack tip.}
\label{stickslip}
\end{figure}

Consider an elastic solid sliding on a flat rigid substrate.
Assume that the elastic solid occupies the space $0<y<H$, and let
$(x,y,z)$ be a coordinate system with the  plane $y=0$ corresponding to
the surface  of the solid, see Fig. \ref{stickslip}. 
We discuss the plane strain
situation with $u_z=0$, where $\bf u$ is the displacement vector and static
elasticity since all  motions are expected to be slow compared to the sound
velocities.

Let us discuss the boundary conditions. 
First of all we assume that displacements at the top boundary of the block
are the same as for the top of the rigid plate: $u_y=0$ and $u_x=vt+ const$ 
are uniform for $y=H$. Here $v$ is the driving velocity of the upper 
rigid plate relative
to the bottom rigid substrate; $t$ is time.    
We assume also that the bottom interface can be in two states: "stick" and
"slip". The boundaries between these two states are described by the
crack edges which move with a  velocity $c$ in the  $x$
direction. In the stick regions $u_y=0$ and
the sliding velocity  also vanishes, $\dot u_x=0$ for  $y=0$.
In the stationary situation, when all stick-slip boundaries
drift with the same velocity $c$, displacement derivatives depend 
on $x$ and time $t$ only in the combination $x-ct$.
Therefore, in the stick regions there is a constant deformation 
$u_{xx}=V_0\equiv v/c$.
 
In the slip regions
we assume that the two solids are always in contact,
$u_y=0$ for  $y=0$, while we allow for a finite
relative sliding velocity $\dot u_x$. This sliding velocity leads
to  frictional shear stress at the interface,
 $\sigma_{xy}=\sigma_{xy}\{\dot u_x(t,x)\}$. In the following, a simple
 linear viscous friction law will be considered,
$
\sigma _{xy}=\beta \dot u_x,
$
with $\beta$ being the  friction coefficient.

Let us turn now to thermodynamical aspects of the problem
Since the adhesion contact in the stick region is stronger,
it is reasonable to assume that the interface energy in
the stick phase is smaller than the interface energy in the slip
phase. We denote this energy difference by $\gamma$. It is clear that
without external loading the stick phase is energetically
favorable and a finite shear stress is required to get the interface
into the slip state.

We first discuss the situation where the external shear stress 
$\sigma_{xy}$ is given
(not the driving velocity). Then we have two homogeneous solutions to 
our problem: i) uniform stick with pure elastic response and no drift 
velocity and ii) uniform sliding with drift velocity 
\begin{equation}
v=S\sigma_{xy}/\mu,
\label{sliding}
\end{equation}
where $\mu$ is the shear modulus and $S=\mu/\beta$ is the velocity scale given
by the friction law. 
We note that this homogeneous sliding mode is
linearly stable for any velocity with respect to small
inhomogeneous perturbations of the stress and strain fields. 
In this respect the viscous
friction law is very different from Coulomb friction which leads
to a linear instability and ill-posedness of the problem as 
has been intensively discussed in  the literature \cite{Ranjith01}.

In our previous paper we
considered a body uniformly strained with $u^\infty_{xy}$ at $x\to\infty$ 
and relaxed at $x\to-\infty$ due to the presence of the slip at $x<0$ 
(a singular solution of the static theory of elasticity similar to a 
crack solution, see Fig. \ref{stickslip}).
The condition for the boundary between semi-infinite slip and stick regions 
to propagate in the positive
$x$-direction is that the energy release because of the stress relaxation is
larger than the energy flux to the crack tip (which is proportional to the 
surface energy): 
\begin{eqnarray}
\Delta=2\mu \left(u^\infty_{xy} \right)^2 H/\gamma >1, 
\end{eqnarray}
where $H$ is the height of the elastic block.
Otherwise, the crack would propagate with negative
velocity and the stick phase would be restored. 
Equilibrium corresponds to
\begin{eqnarray}
\sigma^\infty_{xy}=\sigma_c\equiv\sqrt{2\mu\gamma/H}.
\label{sigmac}
\end{eqnarray}
The condition
$\Delta=1$ or $\sigma_{xy}=\sigma_c$ is nothing but the usual 
Griffith threshold for crack
propagation. On the other hand, in the context of the friction
problem, this condition may be interpreted as a static friction
threshold: a finite shear loading is required to get the system into the
sliding mode.
The energy released by the creation of a single 
slip of length $l \gg H/\sqrt{1-2\nu}$
in the body, uniformly stressed with $u_{xy}$, is 
$(2\mu \left(u_{xy}\right)^2 H -\gamma)l$.
Uniform stick in the long sample is stable if the 
applied shear stress does 
not exceed  $\sigma_c$, otherwise stick is only metastable.
We note that for soft materials such as gels the characteristic length 
$H/\sqrt{1-2\nu}$ can be quite large.  For  length scale, shorter
than this length,  $\sigma_c$ increases  by factor of 
2 compared to Eq. (\ref{sigmac}). We will return to this point later when 
accurately solving the elastic problem in Appendix \ref{incompress}.

\begin{figure}[h]
\includegraphics[width=0.8\linewidth]{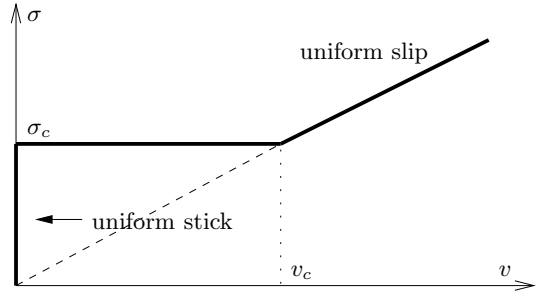}
\caption{Two homogeneous regimes, stick and slip;
$\sigma_c$ is the transition point and also the average stress 
in stick-slip regime. }
\label{phasediag}
\end{figure}

On the other hand, the homogeneous sliding mode may be 
unstable against a resticking pulse 
(nonlinear ``healing instability'') if the
corresponding value of $\sigma_{xy}< \sigma_c$. 
Since in this case the shear stress  is related to the
steady-state sliding velocity by Eq. (\ref{sliding}), we presume that the
homogeneous sliding is stable against the healing instability only
above the critical sliding velocity $v_c$,
\begin{equation}
v_c=S\sigma_c/\mu.
\label{vc}
\end{equation}
Thus, for given $\sigma$ we discuss a dynamical first order phase 
transition between homogeneous stick and 
slip regimes with $\sigma_c$ being the transition point 
(see Fig. \ref{phasediag}).
\psfrag{sigma}{$\sigma$}\psfrag{sigma_c}{$\sigma_c$}
\psfrag{v}{$v$}\psfrag{v_c}{$v_c$}
\psfrag{uniform slip}{uniform slip}
\psfrag{uniform stick}{uniform stick}
Another example of such a non-equilibrium transition is a dielectric 
breakdown above some critical value of the applied electric 
field. The chain reaction of the ionization process takes place 
forming a conducting state instead of a nonconducting one  
realized below the critical value of the applied electrical fields. 
We can also mention a phenomenon of shear banding in complex fluids.
Note that these transitions are not real thermodynamical transitions 
since the slip phase (or conducting state)
is not a real thermodynamical phase and dissipation
due to friction takes place in this regime. Therefore, our arguments 
that $\sigma_c$ is a transition point are not so strict as in usual 
thermodynamics. However, we believe that this picture is plausible and 
also in the spirit  of ``shear melting'' in ultrathin
lubricant layers, discussed in \cite{Aranson02}. In our case, however,
the surface properties are strongly coupled to the bulk elasticity and 
the critical stress $\sigma_c$ depends not only on the interface energy
difference, $\gamma$, but also on geometry. This is the usual situation 
in fracture problems and, as we have already mentioned,  $\sigma_c$ is
just the Griffith threshold for crack propagation in this context.

If one starts from the stick phase and increases the stress slightly 
above the critical value, the transition to the slip regime goes via 
a nucleation process of slip regions and their growth.  Eventually 
annihilation of the boundaries brings the system into the homogeneous 
sliding mode. If the sample is not too long, propagation 
of just one crack may be sufficient to bring the system into the sliding 
mode. The velocity $c$ of such an isolated crack has been calculated in our 
previous paper \cite{Brener02}: 
\begin{eqnarray}
\label{vtip}
c=\frac{\pi\ln \Delta}{\varpi\ln(H/a)}S.
\end{eqnarray}
Here $\varpi$ is a number related to the  Poisson ratio $\nu$:
\begin{eqnarray*}
\varpi=\frac{3-4\nu}{2(1-\nu)}\,,
\end{eqnarray*}
and $a$ is the small length scale cutoff (we will return to this point later).  

If one starts from the sliding mode and reduces the stress below 
the critical value, the transition to the stick phase is via 
nucleation and growth of the resticking regions. 

We note that, for a given shear stress, 
this picture  does not allow for a steady periodic 
stick-slip motion. 
The transition between homogeneous stick and slip states
 corresponds to the critical value of the stress but, 
of course, there may be some hysteresis  around the transition 
point, $\sigma_c$. 

The situation changes drastically when, instead of the shear stress, 
the driving velocity is prescribed. Actually,
 most of the experimental setups fix the driving velocity $v$.
In this case the uniform stick state is no longer possible.  On the 
other hand, if the driving velocity is below its critical value, given by 
Eq. (\ref{vc}),  uniform sliding may be unstable against  resticking 
pulses (nonlinear ``healing instability''). Then we expect the system to
exhibit an inhomogeneous sliding mode, namely periodic 
(quasiperiodic or or even chaotic) stick-slip. 

At this point the 
analogy to the first order phase transitions can be made even deeper:
In thermodynamics, if  the volume but not the pressure  
of the system is prescribed  
for a given temperature,  phase coexistence occurs 
in a range of volumes. If  the volume is changed, the pressure
remains constant according to the equilibrium phase diagram and only 
the relative fraction of the two phases changes. If the volume reaches its
critical value, one phase disappears and phase equilibrium is 
not possible for further volume changes. 
Of course, the driving velocity, 
which is a dynamical variable, cannot be mapped directly onto 
the thermodynamical variable, the volume, as well as the 
hydrostatic pressure is not the same as the shear stress.     
However, we use this  plausible  analogy
and  assume that stick-slip motion  (coexistence of two phases) will 
occur in a range of driving velocities, $v<v_c$, with the average
shear stress  $\sigma\approx\sigma_c$, as shown in 
Fig. \ref{phasediag}.   

Our main goal 
is to discuss this inhomogeneous mode 
of sliding.  In this paper we analyze this problem 
only for the steady-state periodic stick-slip motion schematically
presented in Fig. \ref{stickslip}. The first step is to solve the 
elastic problem for this geometry and 
then to calculate the energy flux into the crack tip and 
the resticking end which allows to formulate equations of motion for these
boundaries. This rather technical issue is discussed in detail 
in the next section. But before some remarks are in order.

The solution of the elastic problems defines a family of the 
displacement fields and slip velocity fields, each of which is labeled
by five parameters: driving velocity $v$, average shear stress $\sigma$,
crack velocity $c$, two length scales of the slip and 
stick regions, $\lambda_{sl}$ and  $\lambda_{st}$ (or alternatively,
the wavelength of the pattern $\lambda= \lambda_{sl}+\lambda_{st}$
and the relative fraction of the slip $\eta=\lambda_{sl}/\lambda$).
The driving velocity $v$ is imposed. This defines a problem of a 
dynamical selection, namely, if a sliding pattern exists, are $c$,
$\lambda_{sl}$, $\lambda_{st}$ and $\sigma$ uniquely defined when 
$v$ is fixed, or not? To find these four parameters we expect to 
have only two equations of motion for the crack tip and resticking end.
Thus, our steady-state periodic problem has a high degeneracy. 
One needs two additional equations to predict a uniquely 
selected pattern. Using the analogy to the first order phase 
transitions we have already removed one degree of freedom, selecting 
the average shear stress $\sigma$ to be close to its critical value 
$\sigma_c$. However, as in many pattern forming systems the selection
of the primary wavelength $\lambda$ is a highly nontrivial issue.
In   directional solidification and eutectic growth (see, for 
example, \cite{Kurz2001}) it is believed that there is only a
``soft'' selection mechanism which depends on the initial conditions,
the history of the process, the level and origin of fluctuations, 
details of the nucleation and annihilation processes. A detailed
discussion of this issue is far beyond the scope of this paper, but 
we  use an additional assumption which  leads
to a ``soft'' selection of the wavelength:  we assume that    
the smallest length scale among $\lambda_{sl}$ and $\lambda_{st}$ is 
 of the order of the sample height $H$. 

\section{Solution of the elastic problem and equations of 
motion for cracks.}
\label{sec:3}
In this Section we solve the problem of linear elasticity with 
the boundary conditions specified above.
We find the expressions for the energy fluxes into the crack edges 
in terms of the imposed boundary conditions (driving velocity $v$, average
shear stress $\tau^*=\sigma$, crack tip velocity $c$) 
and geometrical parameters (sample height $H$, 
pattern wavelength $\lambda=2\pi/k$ and slip fraction 
$\lambda_{sl}/\lambda=\eta$).
We restrict our attention to the very slow modes assuming that crack 
velocities are much smaller than the shear wave speed. 
Stationarity of the problem shows in the fact that the spatial coordinate
$x$ enters only in combination with time: $x-ct$. To simplify
formulas,  we will usually omit the time dependence in the combination.

In Appendix \ref{general} we consider stationary solutions of the linear
elasticity theory in the given geometry and
with boundary conditions $u_y|_{y=0}=u_y|_{y=H}=\partial_x u_x|_{y=H}=0$.
We treat a so called mixed problem of elasticity: there are parts of the
bottom surface (slip regions) where only a linear relation between shear 
strain and stress is given. In the particular case of zero friction,
shear stress is zero.
 
Using the solutions, we can express any relevant quantity (strain or stress
components) in terms of the displacement $u_x|_{y=0}$, or equivalently
in terms of the dimensionless sliding velocity 
$V\equiv \dot u_x/c =V_0-\partial_x u_{x}|_{y=0}$. The relations are generally
nonlocal, leading to an integral equation for the function $V$.  

\subsection{Small wavelengths, $\lambda <<H$.}

Let us consider the case where the height is the largest geometrical
parameter.
Then, as in the case of infinite height,
we obtain the following integral equation \cite{Caroli2000}
\begin{align}
\tau(\xi)=\tau^{*}+\frac{2\mu}{\pi\varpi}\,
{\rm P.V.}\int\limits^{d}_{-d}d\zeta\frac{1+\xi\zeta}{\xi-\zeta}\Phi(\zeta)\,,
\label{tau}  
\end{align}
where $\tau$ is the shear stress at $y=0$ and 
\begin{eqnarray*}
\varpi=\frac{3-4\nu}{2(1-\nu)}\,.
\end{eqnarray*}
Here
we made the variable transformation
$\xi=\tan(kx/2)$, which projects a period $-\pi<kx<\pi$
onto the infinite interval $-\infty<\xi<\infty$, and defined a new function 
$\Phi(\xi)=V(2\arctan \xi)/(1+\xi^2)$ which is nonzero in the slip  
region $|\xi|<d\equiv \tan(\alpha/2)$. The average dimensionless velocity
is given by $V_0=v/c$.

First we consider the solutions of the equation in the absence of friction
in the slip region $k|x|<\alpha$,
which means $\tau(\xi)=0$ for $|\xi|<d$  
($\eta=\alpha/\pi=\lambda_{sl}/\lambda$ is the fraction of the slip). 
The solutions were discussed in Ref. \cite{Caroli2000} (even with a more
complicated form of the viscoelastic bulk response):

\begin{eqnarray}
\Phi_0(\xi)=\frac{\sqrt{1+d^2}}{2\mu}
\frac{\varpi\tau^{*}\xi+2\mu V_{0}}{(1+\xi^2)\sqrt{(d^2-\xi^2)}}\,.
\label{Phi0}
\end{eqnarray}
The solution (\ref{Phi0}) can be rewritten for the strain:
\begin{align}
\partial_x u_x|_{y=0}=V_0-
\frac{\varpi\tau^{*}\tan(kx/2)+2\mu V_{0}}
{2\mu\cos(\alpha/2)\sqrt{\tan^2(\alpha/2)-\tan^2(kx/2)}}
\end{align} 
in the slip region $|\tan(kx/2)|<\tan(\alpha/2)$, and 
$\partial_x u_x|_{y=0}=V_0$ in the stick region.
The singularities in the points $|x|=\alpha/k$ are similar
to those in a crack solution in $2d$ elasticity theory.
Generally, there is a macroscopic elastic energy flow to a point 
with the square-root singularity.

The energy flow is determined by the angular dependence of 
the solution in the vicinity of the point. Thus, we can apply the 
result for the semi-infinite slip \cite{Brener02} to find the energy
flows to the head tip $J_1$ and from the rear tip $J_2$ of the slip:
\begin{eqnarray}
J_{1,2}=
\frac{\pi c(2\mu V_0 \pm \varpi \tau^*\tan(\alpha/2))^2}
{4\varpi\mu k \tan(\alpha/2)}\,.
\end{eqnarray}  
We see that $J_2$ vanishes only if $2\mu V_0-\varpi \tau^*\tan(\alpha/2)=0$,
i.e. when the sliding velocity tends to zero near the rear tip.
This special solution was discussed in the literature 
(see, for example, Refs. \cite{Caroli2000,n00,f79} and references therein).

The energy flowing to the tips can be reversibly converted into the interface 
energy and can be dissipated:
\begin{eqnarray}
J_{1}=\gamma c(1+c/c_0), \qquad J_{2}=\gamma c(1-c/c_0),
\label{dis}
\end{eqnarray}
where $c_0$ describes tip dissipation. 
Let us assume that the tip velocity is so small that the dissipation  
can neglected. 
The resulting conditions, 
$$J_{1}= J_{2}=\gamma c$$  
are 
fulfilled  
if either $V_0=\sqrt{\varpi \gamma k \tan(\alpha/2)/\mu}$ and $\tau^*=0$,
or $\tau^*=2\sqrt{\gamma \mu k/(\varpi\tan(\alpha/2))}$ and $V_0=0$.
The latter case corresponds to the periodic array of slip regions
moving with velocity $c$; in the co-moving frame the strain distribution 
is the same as in the statics. This solution is  unstable. For example,
for the fixed wave number $k$ and  stress smaller than its equilibrium
value, $\alpha$ will decrease until the stick conditions hold everywhere.
The former solution has the feature that the sliding preserves its direction
($V>0$) in the whole slip region.

The case with non-zero friction can be treated as above.
Now, in the left-hand side of Eq. (\ref{tau}), in the 
slip region $|\xi|<d$, we have $\tau(\xi)=\mu c (1+\xi^2) \Phi(\xi)/S$.
The solution can be guessed from that for the semi-infinite slip
\cite{Brener02}:
\begin{eqnarray}
\Phi(\xi)= \frac{b_1\xi+b_2}
{(1+\xi^2)(\xi+d)^{(1+\epsilon)/2}(d-\xi)^{(1-\epsilon)/2}}\,,
\label{fieps}
\end{eqnarray}
where $b_1$, $b_2$ and $\epsilon$ are given by the following expressions:
\begin{align}
\label{b1}
&
b_1=\frac{\cos\frac{\pi\epsilon}{2}}{2\mu\cos(\alpha/2)}
\left(-2\mu V_0\sin\frac{\epsilon\alpha}{2}
+\varpi\tau^*\cos\frac{\epsilon\alpha}{2}\right)\,,
\\
\label{b2}
&
b_2=\frac{\cos\frac{\pi\epsilon}{2}}{2\mu\cos(\alpha/2)}
\left(2\mu V_0\cos\frac{\epsilon\alpha}{2}
+\varpi\tau^*\sin\frac{\epsilon\alpha}{2}\right)\,,
\\
&
\tan\frac{\pi\epsilon}{2} = \frac{\varpi c}{2S}.
\end{align}

Elastic energy flow in the presence of friction $\sigma_{xy}\propto \dot u_x$
depends \cite{Brener02} on the distance from the tip $r$:
\begin{eqnarray}
J_1= \frac{\mu c^2}{S\epsilon} K^2  r^\epsilon\,, 
\end{eqnarray}
where $K$ is a coefficient in front of the main contribution
in the strain $|\partial_x u_x|=K|r|^{(-1+\epsilon)/2}$ near the tip in 
the slip region at $y=0$. In the expression for the rear tip $\epsilon$
is replaced by $-\epsilon$.
It can be easily seen that on each scale the change of the elastic 
energy flow equals the energy loss due to friction.
Thus, we must allow that on the microscopic
scale $a$ the friction law is changed so that the elastic 
energy flow on smaller scales remains constant, the same as for $r=a$.
For example, this should be the case, 
if the frictional shear stress saturates at large velocities. 
We assume that this microscopic cutoff $a$ is a material property
which can be considered to be the same for the head and rear tips
and essentially is independent of $\epsilon$. Then we 
can express the energy flows in terms of $V_0$, $\tau^*$ and $\epsilon$:
\begin{align}
\label{fluxes}
&
J_1= \frac{c
\left(2\mu V_0\cos\frac{(1+\epsilon)\alpha}{2}+
\varpi\tau^*\sin\frac{(1+\epsilon)\alpha}{2}\right)^2 (ka)^\epsilon}
{(2\sin\alpha)^{1+\epsilon}\varpi\mu k }
\frac{\sin(\pi\epsilon)}{\epsilon}, 
\\ 
&
J_2= \frac{c
\left(2\mu V_0\cos\frac{(1-\epsilon)\alpha}{2}-
\varpi\tau^*\sin\frac{(1-\epsilon)\alpha}{2}\right)^2 (ka)^{-\epsilon}
}
{(2\sin\alpha)^{1-\epsilon}\varpi\mu k }
\frac{\sin(\pi\epsilon)}{\epsilon}.
\nonumber
\end{align}

We argue that if the height is larger than either of the lengths 
$\lambda_{sl}$  or $\lambda_{sl}$ and smaller than the other one, 
the solutions for large height are valid essentially. 
Minor changes can be determined with the help of the solutions for 
isolated stick and slip regions given in Appendix \ref{isolated}.

\subsection{Large wavelengths, $\lambda >> H$.}
\label{subsec:3.2}
In the limit of long slip and stick regions, 
one obtains ``zero-mode''
solutions with the relaxed stresses modified by the presence of friction. 
The solutions are valid far from the slip edges. 
In the stick region $|x|<\lambda_{st}/2$,  

\begin{align}
&
u_x = V_0 \frac{(H-y)x}{H}+\frac{\sigma_0 y}{\mu}\,,\qquad
u_y=V_0 \frac{y(y-H)}{4(1-\nu)H}\,,\\
&
\sigma_{xy}=\sigma_0-\mu\frac{x}{H}V_0\,,
\end{align}
and the normal stresses $\sim \mu V_0/(1-2\nu)$ depend only on $y$.
Here $\sigma_0$ is related to the displacement $u_0$ of the sample top with
respect to the middle of the stick: $\sigma_0=\mu u_0/H$.

In the slip region shear stress is relaxed to $\mu v/S$ and 
the sliding velocity  
is equal to the driving velocity $v$ 
with  exponentially small corrections far from the boundary points.

Since, in the limit of small height, 
the normal stresses are relaxed in the slip region, 
we can find the energy fluxes  similarly to the case of  
semi-infinite slip \cite{Brener02}.
 
Using the asymptotic solutions described above, we get energy
flows to (from) the regions of length  
$\sim H$ surrounding boundary points:
\begin{eqnarray}
J_{1,2}^{(0)}=\frac{\mu c H}{2} \left( \frac{\sigma_0}{\mu} 
\pm\frac{\lambda_{st}}{2H}V_0 \right)^2\,,
\end{eqnarray}
where the subscripts $1$ (upper sign) and $2$ (lower sign) correspond to the 
regions near the rear and head tips of the stick region, respectively. 

Without friction, inverse-square-root singularities appear in the 
strains and stresses within a distance $\sim H$ around the boundary points.
The friction changes the exponents to $-1/2\pm\epsilon$ and creates additional 
dissipative energy flows from the sample. These dissipative flows are also 
restricted within the distance $\sim H$ in the slip regions.

Thus, the expressions for the energy flows, instead of Eq. (\ref{fluxes}),
take the following form in the limit of small height:
\begin{align}
\label{fluxesH}
&J_1=\frac{\mu c H}{2} \left( \frac{\sigma_0}{\mu} 
+\frac{\lambda_{st}}{2H}V_0 \right)^2 \left(\frac{H}{a}\right)^{-\epsilon}
\,,\\
&J_2=\frac{\mu c H}{2} \left( \frac{\sigma_0}{\mu} 
-\frac{\lambda_{st}}{2H}V_0 \right)^2 \left(\frac{H}{a}\right)^{\epsilon}\,.
\nonumber
\end{align}
These equations can be rewritten in terms of the average shear stress $\tau^*$ 
using the relation
\begin{align}
\sigma_0=\frac{\lambda}{\lambda_{st}}\left(\tau^*-\mu\frac{v}{S}\right)\,.
\end{align}

Strictly speaking, if $1-2\nu\ll 1$, the relations derived here are 
not valid on intermediate scales 
$H\ll \lambda_{st},\,\lambda_{st}\ll H/\sqrt{1-2\nu}$. 
In Appendix \ref{incompress} we show how to treat the elastic problem
on all scales larger than $H$ without friction. The results on intermediate 
scales do not substantially differ from those obtained above 
(only by numerical factors if written in terms of $\tau^*$).

\subsection{Stability and degeneracy}
Solving the elastic problem we have found energy flows 
to the boundary points of slip pulses $J_{1,2}$.
Equations of motion for the tip and the resticking end of the crack 
are given by the energy conservation, Eqs. (\ref{dis}).  
Therefore, we cannot determine four quantities
($c$, $\tau^*$, $k=2\pi/\lambda$, 
$\eta=\alpha/\pi=(\lambda-\lambda_{st})/\lambda$) using only two equations.
Two of the parameters cannot be chosen directly, and we are faced
with the problem of  dynamical selection. 
We will address this issue in the next section
using stability arguments.

There is another kind of degeneracy related to the fact that the sign of the
term in the parentheses in Eqs. (\ref{fluxes}) and (\ref{fluxesH}) for $J_2$
can be either negative or positive, depending on relative strength of the
average shear stress and  velocity. For simplicity, we consider 
this question in the case of large wavelengths. For small wavelengths, 
the conclusion will be similar.
In the following we neglect 
the tip dissipation ($c/c_0 << 1$) assuming that the main dissipation comes 
from the friction in the slip region. Then Eqs. (\ref{dis}) reduce to 
$J_1=J_2=\gamma c$.
The velocities of the rear tip ($c_1$) and the head tip ($c_2$) of 
the stick region are
(see Eq. (\ref{vtip}) and (\ref{fluxesH}))
\begin{eqnarray}
c_1= \frac{\pi S \ln \Delta_1}{\varpi\ln(H/a)}\,,
\qquad
c_2= -\frac{\pi S \ln \Delta_2}{\varpi\ln(H/a)}\,,
\label{c1c2}
\end{eqnarray}  
 where 
\begin{eqnarray}
\Delta_{1,2}=\frac{H}{2\mu\gamma} 
\left(\frac{\mu V_0 \lambda_{st}}{2H} \pm \sigma_0 \right)^2\,.
\end{eqnarray} 
For the steady-state motion $c_1=c_2=c$ and $V_0=v/c$.
There are two branches of the solution.
On the first branch, we have:
\begin{eqnarray}
\left(\frac{\mu V_0 \lambda_{st}}{2H} \right)^2-\sigma_0^2=\sigma_c^2\,,
\label{1stbranch}
\end{eqnarray}
thus $\mu V_0 \lambda_{st}/(2H)\ge \sigma_c$, and
$\sigma_0$ can be arbitrarily small.

On the second branch,
\begin{eqnarray}
\sigma_0^2-\left(\frac{\mu V_0 \lambda_{st}}{2H} \right)^2=\sigma_c^2\,.
\end{eqnarray}
This branch contains the special static solution $c=v=0$, $V_0=0$ for 
$\sigma_0=\sigma_c$. 
This solution is unstable 
with respect to a change of $\lambda_{st}$.
The instability can be seen from the fact that the
equilibrium corresponds to the energy maximum. 
On the contrary, for the special static solution with $\sigma_0=0, c=0$ 
on the first branch, the total energy of the stick 
is dominated for small $\lambda_{st}$ by the surface energy 
$-\gamma \lambda_{st}$, and for large $\lambda_{st}$ by the elastic
energy 
\begin{eqnarray}
H\int\limits_{-\lambda_{st}/2}^{\lambda_{st}/2} dx\,\frac{1}{2\mu}
\left(\frac{\mu V_0 x}{H} \right)^2 =
\frac{\mu V_0^2 \lambda_{st}^3}{24H}\,,
\end{eqnarray}
and hence there must be a minimum of the total energy corresponding
to the stable equilibrium.

A more general stability criterion  can be derived
as follows. Consider the effect of a small change of the stick length 
$\delta\lambda_{st}$. We suppose that resticking occurs with the same 
strain $V_0$. 
The change of the stick length leads to the changes of the tip velocities: 
\begin{eqnarray}
\delta c_1 = \frac{2\pi S }{\varpi\ln(H/a)}\, 
\frac{\mu V_0}{\mu V_0 \lambda_{st}  + 2H\sigma_0}\delta\lambda_{st} \,,\\
\delta c_2 = -\frac{2\pi S }{\varpi\ln(H/a)}\, 
\frac{\mu V_0}{\mu V_0 \lambda_{st} - 2H\sigma_0}\delta\lambda_{st} \,,
\end{eqnarray}
and hence,
\begin{eqnarray}
\frac{\delta c_2-\delta c_1}{\delta\lambda_{st}}=
- \frac{2\pi S }{\varpi\ln(H/a)}\,
\frac{2(\mu V_0)^2\lambda_{st}}{(\mu V_0 \lambda_{st})^2  - (2H\sigma_0)^2}\,.
\end{eqnarray}
The necessary stability condition 
$(\delta c_2-\delta c_1)/\delta\lambda_{st}<0$ is 
satisfied only on the first branch.

\section{Dynamical selection of the stick-slip pattern}
\label{sec:4}
We note that no deterministic approach can fix the two free parameters
which appear in the result. In this section, 
we discuss the selection problem for 
the parameters of the stick-slip  pattern on the example of a long sample,
starting from the limit of small driving velocities.
 
We focus our attention on the first stable branch. 
For  small heights, 
different stick regions do not interact, and each of them is
characterized by its length $\lambda_{st}$, propagation velocity $c$ and
shear stress parameter $\sigma_0$ (see Eqs. (\ref{c1c2}-\ref{1stbranch})):
\begin{eqnarray}
\left(\frac{\mu V_0 \lambda_{st}}{2H} \right)^2-\sigma_0^2=\sigma_c^2\,,
\qquad 
V_0=\frac{v}{c}\,,
\nonumber
\\
c=\frac{2\pi S }{\ln(H/a)}
\ln\left(\frac{\sqrt{\sigma_c^2+\sigma_0^2}+\sigma_0}{\sigma_c} \right)\,.
\label{longstick}
\end{eqnarray}
In the exactly stationary regime $c$ is the same for different stick 
regions. The solutions ($c,\lambda_{st}$) can be parametrized  by 
$\sigma_0$. The wavelength $\lambda$ is a free parameter.

It is plausible to assume that $\sigma_0$ cannot be much larger than $\sigma_c$.
The inequality $\sigma_0\gg \sigma_c$ would lead to  large values
of the shear stress $\sigma\gg \sigma_c$ in  regions of the length
exceeding $H$. Such a region is unstable with respect to  slip.
Larger values of $\sigma_0$ are not impossible, they are merely less probable.
Thus, we can find an upper estimate for the crack tip velocity $c_{max}$:
\begin{eqnarray}
c_{max}\sim \frac{2\pi S }{\ln(H/a)}\,,
\end{eqnarray}
and a relation for the stick length:
\begin{eqnarray}
%
\lambda_{st}\sim \frac{2Hc\sigma_c}{\mu v}< \lambda_{st,max}
\sim \frac{4\pi HS\sigma_c}{\mu v\ln(H/a)}\,.
\label{st,max}
\end{eqnarray}

We see that the upper limit for $\lambda_{st}$ vanishes
with increasing $v$. In the case of  small heights,  only  
 $\lambda_{st}>H$ is possible and hence  $v<v^*$, where
\begin{eqnarray}
v^* \sim \frac{4\pi S\sigma_c}{\mu \ln(H/a)} = \frac{4\pi v_c}{\ln(H/a)}
\,.
\end{eqnarray}

Concerning the value of $\lambda_{st}$, two cases can be considered.
The first possibility is to have $c\sim c_{max}$ and
$\lambda_{st}\sim \lambda_{st,max}$.

The second possibility following from Eq. (\ref{longstick}) is  
$c\sim c_{max} (\sigma_0/\sigma_c) \ll c_{max}$ 
corresponding  to $\sigma_0\ll\sigma_c$, which would lead to 
\begin{eqnarray}
\lambda_{st} \sim \frac{4\pi HS\sigma_0}{\mu v\ln(H/a)} \ll \lambda_{st,max}
\,.
\end{eqnarray} 
If long stick regions are present in the sample, 
they will eventually catch up with shorter and slower stick regions,
and then probably merge with them. 
Consequently, it is plausible that the  lengths of the stick 
regions are mostly of order of $\lambda_{st,max}$,  
specified in Eq. (\ref{st,max}), and $\tau ^* \sim\sigma_c$.
As $v$ decreases, this length diverges.

On the other hand, the length $\lambda_{sl}$ 
should be of the order of $H$ for small $v$ because  large slip regions,
$ \lambda_{sl}>>H$
with small frictional stress would be unstable against  resticking. 
The size of the critical nucleus of the slip inside the stick phase with
$\sigma_0\sim\sigma_c$ is of the order of $H$ which also prevents 
$\lambda_{sl}$ to be much smaller than $H$.

On the opposite side of the stick-slip regime as $v$ approaches $v_c$, the same 
arguments lead to 
$\lambda_{st}\sim H$. Thus, we can employ the condition
\begin{eqnarray}
{\rm min}\{\lambda_{st},\lambda_{sl}\}\sim H.
\label{cond1}
\end{eqnarray}
This is fully compatible with the condition 
\begin{eqnarray}
\tau^*\sim \sigma_c\,,
\label{cond2}
\end{eqnarray}
which we have found in the small velocity
limit and also using the analogy to the two-phase region in the first order 
phase transitions.

\begin{figure}[h]
\includegraphics[width=1\linewidth]{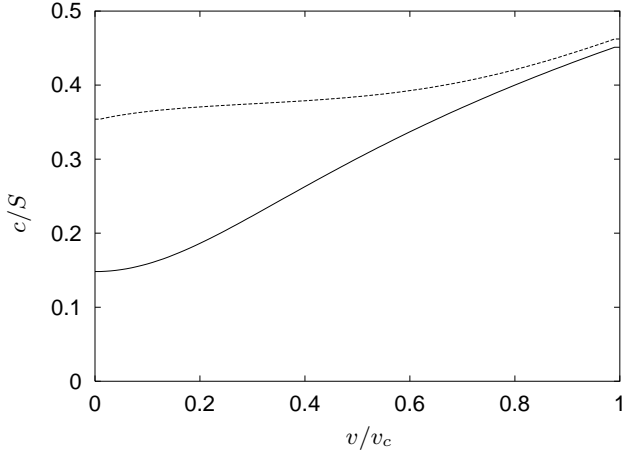}
\caption{Ratio of the crack propagation velocity to the velocity 
scale $S$ set by viscous friction as a function of $v/v_{c}$ for two values 
of $kH/(\pi\sin\alpha)$: 2 (solid line)
and $0.4$ (dashed line);
For all plots $\pi a/(2H)$ was  $2\cdot 10^{-6}$.}
\label{epsilon}
\end{figure}

\begin{figure}[h]
\includegraphics[width=1\linewidth]{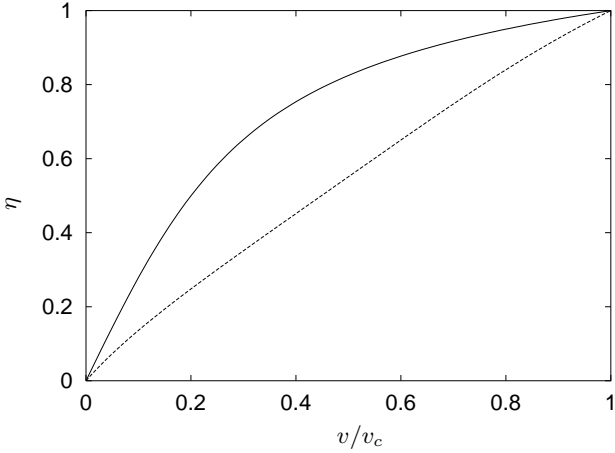}
\caption{Fraction of the slip $\eta=\alpha/\pi$ as a function of 
$v/v_{c}$  for two values of $kH/(\pi\sin\alpha)$: 2 (solid line)
and $0.4$ (dashed line);
$\eta$ does not strongly depend on $kH/(\pi\sin\alpha)$.
For all plots $\pi a/(2H)$ was  $2\cdot 10^{-6}$.}
\label{eta}
\end{figure}

\begin{figure}[h]
\includegraphics[width=1\linewidth]{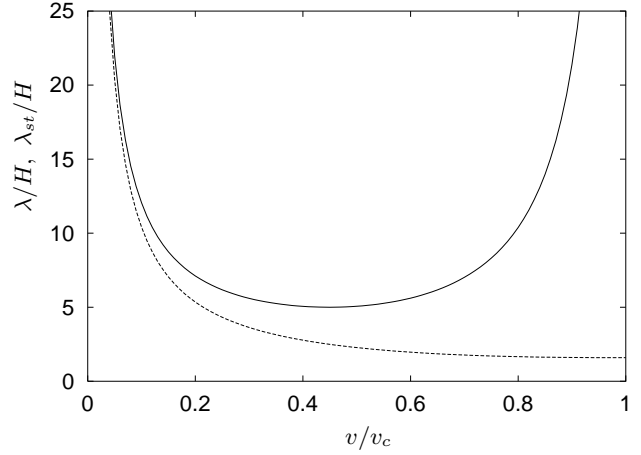}
\caption{$\lambda/H$ (solid line) 
and $\lambda_{st}/H$ (dashed line) as functions of $v/v_{c}$ for  
$kH/(\pi\sin\alpha)=0.4$.
For all plots $\pi a/(2H)$ was  $2\cdot 10^{-6}$.}
\label{lambda}
\end{figure}

The results shown in Figures  \ref{epsilon} - \ref{lambda} represent
numerical solutions of the equations of motion, 
$J_1=\gamma c$ and $J_2=\gamma c$, together with two 
additional constraints (\ref{cond1},\ref{cond2}).
 We model the first condition by
the requirement that $kH/\sin\alpha$ is a fixed number. The dependences
of the slip fraction and the dimensionless crack tip velocities on the 
driving velocity are evaluated both for the limits of large and small heights.
In Fig. \ref{lambda} the values of $\lambda_{st}$ and $\lambda$ are given in 
units of $H$. The crack velocities $c$ remain of the order of $S$ for all 
values of $v/v_c$, the fraction of slip $\eta$ gradually changes from 0 to 1 
and the wavelength $\lambda$ diverges at the both ends of the stick-slip region 
being still of the order of $H$ for the intermediate values of $v/v_c$.

\psfrag{v0}{$v/v_{c}$}
\psfrag{velocity}{$c/S$}

\psfrag{v0}{$v/v_{c}$}
\psfrag{alpha}{$\alpha/\pi$}
\psfrag{eta}{$\eta$}

\psfrag{v0}{$v/v_{c}$}
\psfrag{lam}{$\lambda/H,\ \lambda_{st}/H$}

\psfrag{vsliding}{${\dot u}_x/v_{c}$}
\psfrag{x}{$x/\lambda$}
\begin{figure}[h]
\includegraphics[width=1\linewidth]{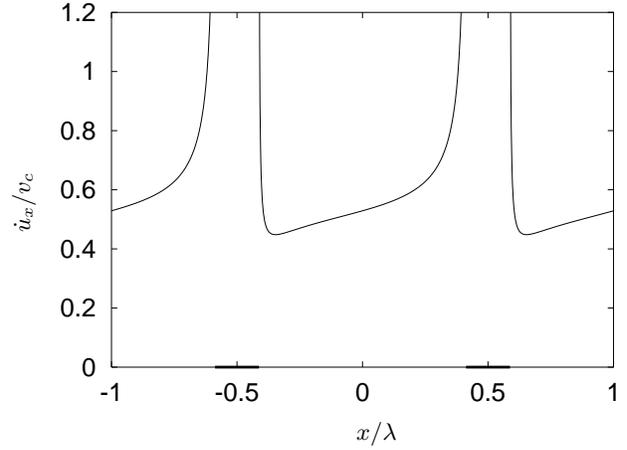}
\caption{Dimensionless sliding velocity ${\dot u}_x|_{y=0}$ 
in the periodic stick-slip regime
as a function of the coordinate 
for $kH/(\pi\sin\alpha)=2$. Two periods are shown. 
The sliding velocity is zero in the stick regions
around $|x|=\lambda/2$. Pattern parameters are calculated for the 
driving velocity $v=0.5v_c$.}
\label{slidingvelocity}
\end{figure}
The sliding velocity in the periodic stick-slip regime calculated from 
Eqs. (\ref{fieps}-\ref{b2}) is presented 
in Fig. \ref{slidingvelocity} for a certain value of the driving velocity.
Note that the singularities at the head and the rear of the slip pulses 
are quite different.

\section{Discussion}
\label{sec:5}

A low velocity dynamics
consisting of periodic slip pulses which heal at a critical
value of the local slip velocity is usually attributed to the 
existence of a V-weakening  frictional regime 
\cite{Baumberger2003}. In this unstable regime the sliding velocity
decreases with the increase of the stress.  
Our model does not contain intrinsic instabilities 
of the friction law as a V-weakening frictional regime.
The homogeneous slip is still metastable below the critical
velocity in our picture.   Critical shear stress and 
critical velocity depend not only on material properties
but also on the geometry and they are related 
to the point of discontinuous
transition (to the Griffith threshold for crack propagation). 
These two scenarios do not contradict  each 
other and can be combined in one model (see, for example, 
\cite{Aranson02} for a description of ``shear induced melting'').
This is  the usual situation in  first order phase transitions:
both, the transition point and instability points exist, forming
the metastable regions around the transition point.
The characteristic velocity scale for the discontinuous transition
and velocity scale for a V-weakening unstable frictional regime can 
be well separated leading to  stick-slip even above the instability
point but, of course, below the transition velocity. 
In our model we discuss the limiting situation where the V-strengthening
regime holds even for very small velocities (at least smaller
than our transition velocity.
  
Now let us discuss the relation of the obtained results to the   
very interesting experimental
observations of Baumberger, Caroli and Ronsin
\cite{Baumberger2002,Baumberger2003}. They performed  experiments of a gel
sliding on a glass plate. The driving velocity was given and the
shear force, and thus the average shear stress, was deduced from the
spring elongation. Above some critical driving velocity
$v_c\approx 100\,\mu$m/s, steady sliding was observed. At 
velocities smaller than the critical one, periodic stick-slip sets in
(see figures in \cite{Baumberger2002,Baumberger2003}). Upon decreasing the
driving velocity $v$, no hysteresis of the transition was
detected. In the stick-slip regime they observed the propagation
of self-healing pulses with no opening,  nucleated  periodically at
the trailing edge of the sample. The propagation velocity of these
cracks was about 60 times larger than the critical sliding
velocity, yet still much smaller than the  shear wave speed.
The crack-like singularities in the slip velocity field were 
detected behind the tip of the pulse but not at the resticking end.

The authors very carefully studied the dependence of the critical
velocity, crack velocities  and other properties of slip pulses 
on the properties of the gel, but the dependence on geometry was not 
discussed. Since both the  critical velocity and the  
critical stress, predicted  by our model,
 depend  on the height of the sample and not only on material
parameters, the crucial test for the relevance of our model to this 
experiment is not possible yet. Moreover, we assumed a very large sample length
compared to its height and discuss a spatially periodic stick-slip pattern,
while in the experiment at most one slip pulse 
was simultaneously observed in the 
sample nucleated at the trailing edge. The length of the slip 
pulse increased with the driving velocity $v$ and sometimes 
the sample was too short to observe more than one stick-slip
boundary. Furthermore,  the fact,  that no hysteresis of the 
transition was detected in the experiment, points towards the 
possible interference of the intrinsic instabilities for low 
sliding velocities, which have  not been  discussed in our model so far.
However, as we have already mentioned, 
the characteristic velocity scale for the discontinuous transition
and the velocity scale for the V-weakening unstable frictional regime can 
be made well separated, if  smaller heights of the sample 
would be available for experimental investigation.

The observed nonlinear behavior of the stress with the
velocity for relatively high sliding rates
(the so-called  shear-thinning rheology) can, in principle, be  
incorporated into a more sophisticated version of the theory presented 
here. This has  already been discussed shortly in the context of 
the microscopic cutoff $a$, the small region around the crack edge 
where the transition from stick to slip takes place.

Another way to approach the observed dependence of the shear stress
on the sliding velocity is to make
a simple modification of the friction law in the slip 
region which amounts to the addition of the constant:
$\sigma_{xy}=const+\beta {\dot u}_x$. This modification does 
not lead to any changes in our results except for a trivial 
shift of all values of the shear stresses.

The predictions of our theory may differ from experimental results
due to different geometries, interference of the intrinsic instabilities,
some specific features caused by polymeric nature of the gel {\it etc}.
The effects on the propagation of single slip pulses are planned
to be considered elsewhere \cite{bmm}.

\begin{figure}[h]
\includegraphics[width=0.9\linewidth]{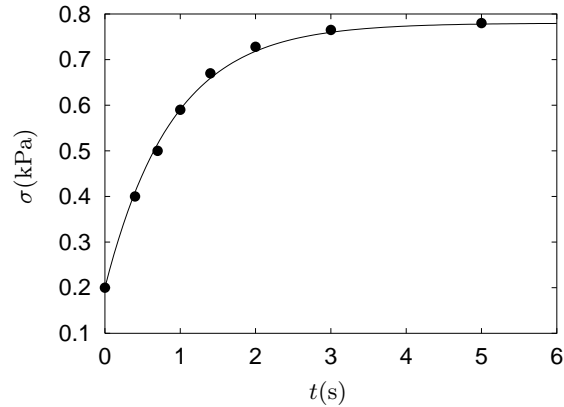}
\caption{Several experimental points taken from Fig. 15 of 
Ref. \cite{Baumberger2003}: relaxation
of shear stress after a jump of the driving velocity.  The 
line represents a fit according to Eq. (\ref{exponential}).}
\label{points}
\end{figure}

Our aim was to construct a generic, conceptually simple model 
which allows for  stick-slip motion in sliding friction without call 
to some very specific properties of materials.    
In our model  all complicated properties
of the gel are hidden in only  two  material parameters which are
related to surface properties: the energy scale $\gamma$ and the 
velocity scale $S$ due to the  viscous frictional law.
Of course, these parameters may depend on the applied normal
stress but we assume that they exist also for zero normal stress.
The velocity scale $S$ can be easily estimated from the shear stress 
response to a jump of the driving velocity in the sliding regime 
(see Fig. 15 in  \cite{Baumberger2003}). This response is 
pure linear in our model and leads to 
\begin{eqnarray}
\sigma(t)=\sigma_{fin}-(\sigma_{fin}-\sigma_{in})\exp(-St/H),
\label{exponential}
\end{eqnarray}
where $\sigma_{in}$ and $\sigma_{fin}$ are the initial and 
final stresses before and after the jump. The response is not changed
if one takes a modified friction law by adding a constant as described
above. 
This exponential behavior is in a good agreement with the experimental 
curve for large changes of stresses and velocities  (see Fig. \ref{points})
and gives an estimate for $S/H\approx 1$ s$^{-1}$, which leads to 
$S\approx 1$ cm/s for used samples with $H=1$ cm.
\psfrag{time}{$t$(s)}
\psfrag{stress}{$\sigma$(kPa)}

One would expect that for ordinary elastic
materials the velocity $S$ should be of the order of the speed of 
sound.
However, for gels the shear modulus $\mu$ is much smaller
than for ordinary materials. The shear sound velocity
$c_t=(\mu/\rho)^{1/2}$ is only $2$ m/s. The velocity
$S=\mu/\beta$ is linear in $\mu$ and should be even smaller.
This is a possible explanation
for a relatively small value of $S$ compared to $c_t$.
  
The existence of a critical sliding velocity, where the transition to a 
stationary sliding occurs, appears
naturally in our description and is given by Eqs. (\ref{sigmac},\ref{vc}).
The length of the slip pulses increase as $v$ approaches $v_c$, and 
the fraction of the slip phase tends to unity in qualitative agreement
with the experiments.  
The characteristic value of the shear strain in the sliding mode near
the critical velocity experimentally was about  $u_{xy}= 0.04$. 
Thus, we can estimate from  Eq. (\ref{sigmac})  the characteristic 
difference between the interface energies in the slip and stick states 
to be $\gamma=0.1 $ J/m$^2$. 

In the stick-slip regime, which exists below $v_c$, the nucleation of
a slip pulse takes place at the trailing edge of the sample and
requires overshooting above the Griffith threshold according to
the experiment. This overshooting is not small and we can use Eq. (\ref{vtip})
 to estimate the crack velocity. Because of the week logarithmic
parameter dependence, we conclude that the crack propagation velocity 
$c$ is of  the order
of $S$ and essentially independent of the driving velocity, in agreement
with experimental observations.  
After the slip pulse traversed the sample the stress drops
below the Griffith threshold and resticking takes place via propagation
of a healing front. Theoretical predictions for the crack velocities
$c$ in the steady-state spatially periodic stick-slip regime are given 
in Fig. \ref{epsilon}: we get $c\sim S\approx 1$ cm/s, 
which is also in qualitative agreement 
with the experiments.

The crack-like singularities in the slip velocity field are predicted 
by our theory and  were detected behind the tip of the pulse.  
At the resticking end the sliding velocity in the experiment 
suddenly drops to zero from a finite
value which is close to the critical velocity. We should note 
that this discontinuity with finite amplitude is not compatible 
with  solutions of the elastic problem which allow either crack-like 
singularities or, as a particular solution,  a smooth, continuous
termination of the slipping process
at the healing point. This type of  solutions,
which correspond to a vanishing stress intensity factor due to some 
additional constraint on the parameters, was discussed in the literature
(e.g. Refs. \cite{n00,f79})
and also in the previous sections. 
Perhaps, one could argue that 
a fast  transient, too fast to be experimentally resolved, takes place
(see Fig. \ref{slidingvelocity}).

The geometry of the  experiment is such that the total
macroscopic friction of the sliding sample 
and the nucleation of the pulses  depend on
the processes  taking place at the edges of the sample. The
stresses here are highly inhomogeneous and the kinetic phenomena
should be considered with a great care.
It would be very interesting to perform a experiment in such 
conditions where the length of the sample and its height are well 
separated  and to study the possible height dependence of the critical 
velocity and the shear stress.
We believe that further theoretical and experimental investigations will 
shed light on this phenomenon where two intriguing problems, crack
propagation and friction, combine together.

\begin{acknowledgement}
We would like to acknowledge fruitful discussions with E.~I.~Kats,
J.~S.~Langer, and V.~Steinberg. We are grateful to H.~Schober for reading
the manuscript and useful comments.
S.M. and V.M. thank the Forschungszentrum J\"ulich
for its hospitality. S.M. thanks 
RFBR for  financial support under grant 
No. 03-02-16173, and DFG for support in the framework of SFB 608.
\end{acknowledgement}

\appendix

\section{Relation between strain and stress at $y=0$}
\label{general}

We  consider only periodic stationary solutions,
possible generalizations are specially mentioned.

A starting point is the momentum conservation 
equation of the linear elasticity theory:
\begin{eqnarray}
\partial_t^2 u_\alpha=(c_l^2-c_t^2)\nabla_\alpha\nabla_\beta u_\beta+
c_t^2\nabla^2 u_\alpha\,,
\end{eqnarray}
where $c_l$ and $c_t$ are the longitudinal and transverse sound velocities
respectively, 
$c_l/c_t=\sqrt{2(1-\nu)/(1-2\nu)}$. 

We consider only the plane strain case with $u_z=0$.
The solutions can be found in form of Fourier series \cite{Caroli2000}:
\begin{align}
u_x=& vt+u_x^*(y)+{\rm Re}\sum\limits_{m=1}^{\infty} \left( 
a_m e^{-ms_-y}+b_m e^{-ms_+y}
\right.\nonumber
\\ 
&
\left.
+c_m e^{ms_-y}+d_m e^{ms_+y}\right)e^{imk(x-ct)},\\
\label{series}
u_y= &{\rm Re} \sum\limits_{m=1}^{\infty} \left( 
a_m \frac{ik}{s_-}e^{-ms_-y} +\frac{is_+}{k}b_m e^{-ms_+y}
\right.\nonumber
\\ 
&
\left.
-\frac{ik}{s_-} c_m e^{ms_-y}
-\frac{is_+}{k}d_m e^{ms_+y}\right)e^{imk(x-ct)},\\
\nonumber
\sigma_{xy}=&\tau^*+\mu{\rm Re} \sum\limits_{m=1}^{+\infty} mk
\left( -a_m \left(\frac{k}{s_-}+\frac{s_-}{k}\right) e^{-ms_-y}
\right.\nonumber
\\ 
&
\left.
-2\frac{s_+}{k}b_m e^{-ms_+y}
+\left(\frac{k}{s_-}+\frac{s_-}{k}\right) c_m e^{ms_-y}
\right.\nonumber
\\ 
&
\left.
+2\frac{s_+}{k}d_m e^{ms_+y}\right)e^{imk(x-ct)} \,,
\end{align}
where
\begin{eqnarray*}
s_+=k\sqrt{1-\frac{c^2}{c_l^2}}, \qquad  
s_-=k\sqrt{1-\frac{c^2}{c_t^2}}\,,
\end{eqnarray*}
and $k$ is the pattern wave number: $k=2\pi/\lambda$.
In a periodic pattern, an average value of a quantity corresponds to its
zero Fourier harmonic. We assume that on average there is no  
normal stress, i.e. $u^*_x(y)$ is a linear function, and there is no
zero-harmonic term in $u_y$.  

We use boundary conditions $u_y|_{y=0}=u_y|_{y=H}=0$ and 
$\partial_x u_x|_{y=H}=0$, 
which give three linear uniform equations
for $a_m$, $b_m$, $c_m$, $d_m$. This allows to express the stress
$\tau\equiv\sigma_{xy}|_{y=0}$ in terms of the Fourier harmonics of the 
dimensionless sliding velocity 
$V\equiv \dot u_x/c =V_0-\partial_x u_{x}$:
\begin{eqnarray*}
\tau(\eta)=\tau^*+\mu{\rm Re} \sum\limits_{m=1}^{\infty}  
(-2i)\chi(m)B_m e^{imk(x-ct)}\,,
\end{eqnarray*}
where $B_m$ is 
\begin{eqnarray*}
&&
B_m=\frac{1}{\pi}\int\limits_{-\pi/k}^{\pi/k}dx\,kV(x)e^{-imkx}\,,
\end{eqnarray*}
and $\chi(m)$ is calculated under the assumption $c^2\ll c_t^2,\ c_l^2$:
\begin{align*}
\chi=2(1-\nu)
\frac{(3-4\nu)\sinh(mkH)\cosh(mkH)-mkH}
{(3-4\nu)^2\sinh^2(mkH)-(mkH)^2}\,.
\end{align*}
This is the result of the static elasticity theory. The first  $c$-dependent
correction would be proportional to $c^2/c^2_t$.

We note that the limit of $\nu\to 1/2$  and $mkH \to 0$ is not well defined:
the limiting value of $\chi$ depends on the ratio $(mkH)^2/(1-2\nu)$.
If $(mkH)^2 \ll (1-2\nu)$ one gets $\chi(m)=1/(2mkH)$. 
The relation between zero harmonics of the shear 
strain and the stress  $\tau=2\mu u_{xy}$ 
is recovered on scales larger than $H/\sqrt{1-2\nu}$.
If $\nu$ is close to $1/2$, there is an intermediate region 
$\sqrt{1-2\nu} \ll mkH \ll 1$ where we have $\chi(m)\approx 2/(mkH)$.
Finally, if $mkH\gg 1$,
\begin{eqnarray*}
&&
\chi=\frac{2(1-\nu)}{3-4\nu}\equiv \frac{1}{\varpi}\,.
\end{eqnarray*}
The latter is realized for the periodic pattern with the wavelength 
smaller than the height, and all the terms in the Fourier series satisfy
the condition $mkH\gg 1$.


\section{Short isolated stick and slip regions}
\label{isolated}
In this Appendix we specify the results for  short isolated slip and stick 
regions. This limit is characterized by the following relation between 
the strain (expressed in terms of dimensionless velocity $V$ in the stationary 
regime) and the stress $\tau(x)\equiv\sigma_{xy}|_{y=0}$:
\begin{eqnarray}
\tau(x)=\frac{2\mu}{\pi\varpi} {\rm P.V.}\int \frac{dz}{x-z}V(z)+\tau^* \,,
\end{eqnarray}
which is derived similarly to the periodic case.
\subsection{Single slip of length $\lambda_{sl}\ll H$}

First consider the case of a short slip, propagating with speed $c$, 
in the presence 
of  friction proportional to sliding velocity; the top surface is uniformly
displaced and fixed (no driving velocity).

The solution with zero friction which satisfies the condition $V=0$ in the
stick region $|x|>\lambda_{sl}/2$ has the following form:
\begin{eqnarray}
V(x)=\left\{
\begin{array}{rcl}
\frac{b_1x+b_2}{\sqrt{(\lambda_{sl}/2)^2-x^2}},& \ \ & |x|<\lambda_{sl}/2\\
0,& \ \ & |x|>\lambda_{sl}/2
\end{array}
\right. \,,
\end{eqnarray}
where
\begin{align*}
&b_1=\frac{\varpi\tau^*}{2\mu}\,,\qquad 
b_2=\frac{1}{\pi}\left(u_2-u_1\right)\,,
\\
&
u_1=u_x|_{x=\lambda_{sl}/2}\,,\qquad u_2=u_x|_{x=-\lambda_{sl}/2}
\,.
\end{align*}

The solution with the linear friction law reads:
\begin{eqnarray*}
V(x)=\left\{
\begin{array}{rcl}
\frac{b_1x+b_2}{(\lambda_{sl}/2-x)^\frac{1-\epsilon}{2} 
(\lambda_{sl}/2+x)^\frac{1+\epsilon}{2}},& 
\ \ & |x|<\lambda_{sl}/2\\
0,& \ \ & |x|>\lambda_{sl}/2
\end{array}
\right. 
\end{eqnarray*}
where
\begin{align*}
&
\tan\frac{\pi\epsilon}{2}=\frac{\varpi c}{2S},\qquad
b_1=\frac{\varpi\tau^*\cos\frac{\pi\epsilon}{2}}{2\mu},
%
\\
&
b_2=\frac{\epsilon\varpi\tau^*\lambda_{sl}\cos\frac{\pi\epsilon}{2}}{4\mu}
+(u_2-u_1)
\frac{\cos\frac{\pi\epsilon}{2}}{\pi}  \,.
\end{align*}

Extracting the coefficients in front of singularities, one 
finds the energy flows:
\begin{align*}
J_{1,2}=
\frac{c\left(2\mu(u_2-u_1)
+\frac{\pi\varpi\tau^*\lambda_{sl}(\epsilon\pm 1)}{2}\right)^2}
{4\pi^2\varpi\mu\lambda_{sl}}
\frac{\sin(\pi\epsilon)}{\epsilon}
\left(\frac{a}{\lambda_{sl}}\right)^{\pm\epsilon}
\!\!\!.
\end{align*}
The energy flows could have been recovered from expressions
(\ref{fluxes}) for periodic stick-slip 
in the limit $\alpha\to 0$, $2\alpha/k=\lambda_{sl}=const$.

\subsection{Single stick of length $\lambda_{st}\ll H$}

The solution with finite friction reads:
\begin{eqnarray*}
V(x)=\left\{
\begin{array}{rcl}
\frac{b_1x+b_2}{(x+\lambda_{st}/2)^\frac{1-\epsilon}{2} 
(x-\lambda_{st}/2)^\frac{1+\epsilon}{2}},& 
\ \ & |x|>\lambda_{st}/2\\
0,& \ \ & |x|<\lambda_{st}/2
\end{array}
\right. 
\end{eqnarray*}
Here
\begin{align*}
&
\tan\frac{\pi\epsilon}{2}=\frac{\varpi c}{2S},\qquad
b_1=V_0=\frac{\varpi\tau^*\cot\frac{\pi\epsilon}{2}}{2\mu},\qquad
\\
&
b_2=
-\frac{\varpi\epsilon\lambda_{st}\tau^*\cot\frac{\pi\epsilon}{2}}{4\mu}
+\left(u_x|_{x=+\infty}-u_x|_{x=-\infty}\right)
\frac{\cot\frac{\pi\epsilon}{2}}{\pi}  \,.
\end{align*}

\section{Small height solutions for almost incompressible media, 
$1-2\nu\ll 1$}
\label{incompress}

The basic idea of the small height approximation is to use the 
approximate equations of elasticity theory which are valid if the 
transverse gradients (along the shortest vertical dimension) 
of displacements are much larger than the longitudinal ones. This is justified 
by the fact that certain boundary conditions are imposed on the upper and lower 
surfaces. The approximation is adequate in the regions 
far from the points of discontinuity in the boundary conditions. 
In those regions, imposing $u_y$ on the boundaries additionally implies
that $\partial_x u_y \ll \partial_y u_x$. The approximation cannot be used
at the distances of the order of $H$ near the crack tips.

Neglecting $\partial^2_x$ terms in comparison with  $\partial^2_y$ terms
in the exact equations, one gets:
\begin{eqnarray}
12\partial_x(\partial_x u_x+\partial_y u_y)+b_0^2\partial^2_y u_x=0\\
12\partial_y(\partial_x u_x+\partial_y u_y)+b_0^2\partial^2_y u_y=0
\label{43}
\,,
\label{simplelast}
\end{eqnarray} 
where $b_0^2=12(1-2\nu)$. The case of $b_0\sim 1$ was discussed in 
Sec. \ref{subsec:3.2}. In the limit $b_0\ll 1$
one can conclude from Eq. (\ref{43}) that 
${\rm div\,} u\equiv \partial_x u_x+\partial_y u_y$ is independent of $y$.
Then, it is easy to see that $u_x$ quadratically depends on $y$,
and $u_y$ is a cubic polynomial in $y$. 

\subsection{Long stick}
In the stick region, we use stationary boundary conditions to
find the solution which is expressed in terms of unknown functions of $x$:
\begin{align}
&{\rm div\,} u=f_1(x)\\
&u_x=y(y-1)g_1(x)+u_0 y-V_0 x(y-1)\\
&u_y=y(y-1)(p_1(x)+yq_1(x))
\,.
\end{align}
Here the lengths are measured in the units of height, and $u_0$ is
the displacement of the upper boundary with respect to the origin. 
Together with the first equation of (\ref{simplelast}),
there are four linear differential equations for the four functions,
which result in $g_1''-b_0^2 g_1=0$. Thus, all functions can be parametrized
by two constants: 
\begin{align}
\label{g1}
&
g_1=a\sinh(b_0x)+b\cosh(b_0x), \quad f_1=\frac{1}{2}V_0-\frac{1}{6}g_1'\,,\\
&
\nonumber
q_1=-\frac{1}{3}g_1', \quad p_1=V_0-f_1=\frac{1}{2}V_0+\frac{1}{6}g_1'
\,.
\end{align}

\subsection{Long slip}
The same procedure can be used for the solution in the slip region.
Here, because of different boundary conditions, $u_x$ has another form:
\begin{eqnarray}
u_x=u_0+\frac{c}{S}V_0(y-1) +(y-1)(h_2(x)+yg_2(x))\,.
\end{eqnarray}  
The presence of the boundary friction (finite $S$) makes this problem
more difficult. The set of equations is reduced to the third-order
differential equation:
\begin{eqnarray}
4h''_2-b_0^2 h_2 + \frac{c}{S}\left( -h'''_2+b_0^2 h'_2\right)=0\,.
\end{eqnarray}  
Without friction the solution can be written as
\begin{align}
&
h_2=A\sinh(b_0x/2)+B\cosh(b_0x/2)
\,.
\end{align}
Small friction changes the balance between the growing and vanishing
exponentials on large scales but the final effect 
can be shown to be negligible.

Functions $g_2$, $p_2$, $q_2$ and $f_2={\rm div\,} u$ 
are expressed in terms of $h_2$, $V_0$ and $c/S$, in analogy to 
Eqs. (\ref{g1}).

\subsection{Matching stick and slip solutions}
The solutions found above are good approximations far from the stick-slip
boundaries. To match the solutions, one must satisfy global mass and momentum
conservations.  At every boundary point this will give two
relations, which equals the number of unknown constants in every interval.

The condition, that the total horizontal force acting on a piece
including the stick-slip boundary is zero, leads to the continuity of
divergences: $f_1(\pm \lambda_{st}/2)=f_2(\mp\lambda_{sl}/2)$.

The second condition is the equality of the volume integral of divergence 
to the surface integral of the displacement. This gives the relations
\begin{eqnarray*}
\pm\frac{1}{4}V_0 \lambda_{st}-\frac{1}{6}g_1(\pm \lambda_{st}/2)
+\frac{2}{3}h_2(\mp\lambda_{sl}/2)=\frac{1}{2}u_0\,.
\end{eqnarray*}
Upper and lower signs correspond to the left and right tips of the slip region.

Deriving the conditions, one neglects $1/\lambda_{st}$ and $b_0$ in 
comparison with unity.
The full set of equations for four constants reads:
\begin{align*}
&
a\sinh(b_0\lambda_{st}/2)+4A\sinh(b_0\lambda_{sl}/4)=\frac{3V_0\lambda_{st}}{2}
\,,
\\
&
b\cosh(b_0\lambda_{st}/2)-4B\cosh(b_0\lambda_{sl}/4)=-3u_0
\,,
\end{align*}
\begin{align*}
&
a\cosh(b_0\lambda_{st}/2)-2A\cosh(b_0\lambda_{sl}/4)=\frac{3V_0}{b_0}
\,,
\\
&
b\sinh(b_0\lambda_{st}/2)+2B\sinh(b_0\lambda_{sl}/4)=0
\,.
\end{align*}

Energy flow to the rear edge of the slip is given by
\begin{align*}
J=&\frac{\mu c}{2}\left(\frac{4}{3}B^2-\frac{1}{3}b^2-u_0^2- 
\frac{2V_0a(\cosh(b_0\lambda_{st}/2)-1)}{b_0} \right.
\\
&\left.
-\frac{2V_0b\sinh(b_0\lambda_{st}/2)}{b_0}+V_0u_0\lambda_{st}
-\frac{1}{4}V_0^2\lambda^2_{st}\right)\,.
\end{align*}
In the limit of not extremely long stick and slip regions,
$H\ll \lambda_{st},\,\lambda_{sl}\ll H/b_0$, the expressions for the 
energy flow are reduced to
\begin{align}
J_{1,2}=\frac{\mu c}{2H}\left(V_0\lambda_{st}
\pm u_0\frac{2(\lambda_{st}+\lambda_{sl})}
{4\lambda_{st}+\lambda_{sl}} \right)^2\,.
\end{align}
Relating the displacement $u_0$ and the average shear stress $\tau^*$ as
\begin{eqnarray*}
u_0=\frac{4\lambda_{st}+\lambda_{sl}}{4\lambda_{st}}\frac{H\tau^*}{\mu}\,,
\end{eqnarray*}
we rewrite the energy flows in the following manner:
\begin{align*}
J_{1,2}=\frac{\mu cH}{2}\left(\frac{V_0\lambda_{st}}{H}
\pm\frac{\tau^*}{\mu}\frac{\lambda}{2\lambda_{st}} \right)^2\,.
\end{align*}
This result differs from the frictionless case of Eqs. (\ref{fluxesH})
only by numerical factors. Moreover, for $\lambda=\lambda_{st}$ and 
$V_0=0$, the energy conservation, $J=\gamma c$, leads to the result 
for the critical stress: $\tau^*=2\sigma_c$.

\end{document}